\documentclass[a4paper,11pt]{article}
\usepackage{pos}
\usepackage{multirow}

\title{Quark mass dependence of hadron resonances}

\author*[a]{F. Gil-Domínguez}
\author[a]{R. Molina}

\affiliation[a]{IFIC (CSIC-UV),\\
  Catedrátic José Beltrán Martinez 2, Valencia, Spain}

\emailAdd{fernando.gil@ific.uv.es}
\emailAdd{raquel.molina@ific.uv.es}

\abstract{We study the dependence of the $D_{s0}(2317)$ mass on the light quark mass through the analysis of data from QCD lattice simulations. Combining HH$\chi$PT and model selection tools as the LASSO method we fit the lattice data of the low-lying charmed mesons masses, obtaining their extrapolation to the physical point and extracting results for the quark mass dependence of the exotic resonances $D_{s0}(2317)$ and $D_{s1}(2460)$, taking as input the energy levels from a lattice simulation.}

\FullConference{%
  The 39th International Symposium on Lattice Field Theory\\
  8–13 August, 2022\\
  Endenicher Allee 19 C, 52115 Bonn, Germany
}

\begin{document}
\maketitle

\section{Introduction}

The discovery of exotic hadrons in the heavy quark sector, which cannot be accommodated in terms of $q\bar{q}$ mesons or $qqq$ baryons, like those with a tetraquark or pentaquark structure, has manifested the relevance of hadronic loops in order to explain the masses and other properties of many states in the hadron spectrum \cite{guoreview}. 
In order to describe the heavy meson spectrum, it is useful to take a look to the heavy quark limit ($m\to\infty$). As it is well-known, the interaction becomes independent on the spin of the heavy quark in this limit due to the inverse color-magnetic-moment dependence on the heavy quark mass. Then, for each $Q\bar{q}$ state there will be another one with the same mass which can be obtained by a flip of the spin of the heavy quark. 
If we focus on the charmed-strange meson spectrum and one assumes that the mass of the charm quark is sufficiently heavy to consider this limit, for $l=1$ one has two doublets. The first one for $J^P_l=\frac{1}{2}^+$, $J^P=0^+,1^+$, might be related to the observed states $D_{s0}(2317)$ and $D_{s1}(2460)$ \cite{babards0,cleods1}. Regarding these states, the constituent quark model predicts broad states decaying to $D^{(*)}K$ for the $J^P=\{0,1\}^+$ doublet with masses higher in about $100$ MeV than the experimental ones \cite{godfrey,godfrey2,dipierro}. In contrast, the observed states, the $D_{s0}(2317)$ and $D_{s1}(2460)$, are very narrow resonances which are placed very close to the $DK$ and $DK^*$ thresholds respectively in the spectrum \cite{babards0,cleods1}. 
For these reasons, molecular \cite{barnesclose,kolomeitsevlutz} and also tetraquark explanations \cite{chenghou,terasaki} have been proposed. 

A new state called $X_0(2866)$ close to the $D^*K^*$ threshold was predicted in \cite{molinabranz,molinaoset} as a molecular state emerging from the hidden-gauge interaction between two vector mesons and recently, the it has been observed in the LHCb \cite{LHCb1,LHCb2}.
Also, a new state close to the $D^*K^*$ threshold, the $T_{c\bar{s}}(2900)$ has been also observed by the LHCb in $B\to \bar{D}D_s\pi$ decays \cite{LHCb:2022xob}.

We can study an strongly interacting system at low energies with a heavy meson $Q\bar{q}$ by means of Heavy Hadron Chiral Perturbation Theory (HH$\chi$PT) \cite{wise,burdman,yancheng}. This effective field theory is based on both chiral and heavy quark symmetries, where a systematic two double expansions in the parameters $Q/\Lambda_\chi$ and $\Lambda_{QCD}/m_Q$ can be made, being $Q$ the momentum transfer (soft scale), $Q\sim m_\pi\sim p_\pi$, $m_Q$, is the mass of the heavy quark, and $\Lambda_\chi=4 \pi f\simeq 1$ GeV (hard scale). 
Given the large amount of parameters in one-loop HH$\chi$PT it has not been possible yet to obtain them precisely. In this work, we will study the quark mass dependence of the ground state charmed mesons $D_{(s)}$ and $D_{(s)}^*$, in the framework of one-loop HH$\chi$PT \cite{jenkins}. This can be possible now since there might be sufficient lattice simulations to better constraint the parameters of HH$\chi$PT. Moreover, there are more advanced statistical techniques which may be helpful in this task like the LASSO regression method.

For this analysis, we have taken as input the lattice data on the ground state charmed mesons of Ref. \cite{kalinowskiwagner}, which uses Wilson twisted mass lattice QCD with 2+1+1 dynamical quark flavors for pion masses in the range of $275-450$ MeV, and a lattice spacing $a\simeq 0.0885$ fm \cite{EuropeanTwistedMassa} [ETMC]; the PACS ensembles with $N_f=2+1$ flavor Clover-Wilson configurations \cite{mohlerwoloshyn} for pion masses in the range $\sim 150-420$ MeV and a lattice spacing $a=0.0907$ MeV \cite{aokiphys} [PACS-CS]; the ensembles from Hadron Spectrum Collaboration for $N_f = 2 + 1$ flavours of dynamical quarks with an anisotropic (clover) action \cite{cheungohara}, see also Table 2 of  \cite{cheungthomas} [HSC];  the data of Table 1 of \cite{prelovsekpadmanath} for a $N_f = 2 + 1$ simulation with Wilson dynamical fermions provided by the Coordinated Lattice Simulations (CLS) for a pion mass of $m_\pi=280$ MeV and a lattice spacing of $a=0.1239$ fm \cite{brunomattia} [CLS]; the data of the simulation with $N_f=2$ improved clover Wilson sea quarks collected in Table 1 of \cite{balicollins} for $m_\pi=150,290$ MeV and $a=0.07$ fm \cite{balicollinsa} [RQCD]. Most of these data are collected in Tables tables VIII-XIV of \cite{guoheo}, where a previous lattice data analysis using the chiral lagrangians developed in \cite{lutzsoyeur,kolomeitsevlutz} was performed. 
Finally, in order to study the quark mass dependence of the $D_{s0}(2317)$ we analyze the lattice data of the HSC energy levels for two different pion masses \cite{cheungthomas}, using the hidden gauge lagrangian which is consistent with the lowest order of HM$\chi$PT \cite{Molina:2010tx,Albaladejo:2018mhb,Gamermann:2006nm}. Previous analysis to $DK$ energy levels lattice data have been done in Refs. \cite{Albaladejo:2018mhb,Lutz:2022enz,MartinezTorres:2014kpc}.

The organization of this work is as follows. First, in section \ref{dmasses} we show the fit of the charmed meson masses to this lattice data. In section \ref{Elvlsfit}, the data of \cite{cheungthomas} are analyzed in the framework of HM$\chi$PT. Finally, the quark mass dependence of the $D_{s0}(2317)$ properties in both, the light and charm quarks is extracted in section \ref{Ds0dependence}.

\section{Charmed meson masses analysis}\label{dmasses}

\subsection{Low-lying charmed meson masses in one loop HH$\chi$PT}

The formulas for the low-lying charmed meson masses ($D$, $D^*$, $D_s$ and $D_s^*$) are calculated at one loop in HH$\chi$PT in the Appendices (A.1) and (A.10) of \cite{jenkins}. These are,

\begin{equation}\label{md1}
\frac{1}{4}(M_{D_a}+3M_{D_a^*})= m_H + \alpha_a-\sum_{X=\pi,K,\eta}\beta_a^{(X)}\frac{M_X^3}{16\pi f^2} +\sum_{X=\pi,K,\eta}\left(\gamma_a^{(X)}-\lambda_a^{(X)}\alpha_a\right)\frac{M_X^2}{16\pi^2 f^2}\log\left(M_X^2/\mu^2\right)+c_a
\end{equation}

and

\begin{equation}\label{md2}
(M_{D_a^*}-M_{D_a})= \Delta+\sum_{X=\pi,K,\eta}\left(\gamma_a^{(X)}-\lambda_a^{(X)}\Delta\right)\frac{M_X^2}{16\pi^2 f^2}\log\left(M_X^2/\mu^2\right)+\delta c_a.
\end{equation}

The meson masses are fitted to the lattice data collected. These formulas depend on the physical masses of the light pseudoscalar mesons and also the terms $\left(\gamma_a^{(X)}-\lambda_a^{(X)}\alpha_a\right)$, $\left(\gamma_a^{(X)}-\lambda_a^{(X)}\Delta\right)$, $c_a$ and $\delta c_a$ depend on light quark masses. These coefficients are given in the Appendices (A.1) and (A.10) of \cite{jenkins}. The light quark masses are related to the pseudoscalar masses through the leading order formulas

\begin{equation}
M_\pi^2=2B_0m, \quad M_K^2=B_0(m+m_s) \quad \rightarrow \quad m=\frac{M_\pi^2}{2B_0}, \qquad m_s=\frac{M_K^2}{B_0}-\frac{M_\pi^2}{2B_0}.
\end{equation}

There are 11 free parameters to determine: ($m_H$, $\frac{\sigma m_\pi}{B_0}$, $\frac{a m_\pi}{B_0}$, $\frac{b m^3_\pi}{B_0^2}$, $\frac{c m^3_\pi}{B_0^2}$, $\frac{d m^3_\pi}{B_0^2}$, $\Delta$, $\frac{\Delta^{(\sigma)} m_\pi}{B_0}$, $\frac{\Delta^{(a)} m_\pi}{B_0}$, $g^2$ and $\mu$). The scale $\mu$ is fixed to $770$ GeV and the coupling constant $g^2$ to $0.55$ to obtain the experimental radiative and pion decay widths of the charmed vector mesons.
The $m_H$ parameter is a power series of $m_Q$. At leading order in this expansion $m_H = m_Q$. The $m_H$ and $\Delta$ parameters can be interpreted as the spin-average mass and hyperfine splitting of the heavy mesons in the $SU(3)$ chiral limit. Then, $m_H$ and $\Delta$ can be different for each lattice collaboration and will be determined for each collaboration independently.
In that way, we end up with $7 +2n$ parameters to fit, with $n$ the number of collaborations, in our case $n=7$.

\subsection{Fitting procedure with LASSO method}

The fit is performed using data from different collaborations and the theoretical formulas of Eqs.(\ref{md1}) and (\ref{md2}).
Thus, as we mentioned, in our case $\vec{p}$ includes the $7+2n$ fit parameters where $m_H$ and $\Delta$ are different for each collaboration. However, in the procedure we have realized that the case of HSC is special, the charmed meson data for the light and heavy pion masses have a slightly different scale setting $a_t$ and charm quark mass determination $a_t M_{\eta_c}$ of the order of the binding energy of the $D_{s0}(2317)$, so we need to separate them as if they were two collaborations, ie HS239 \cite{Cheung:2016bym} and HS391 \cite{HadronSpectrum:2012gic}. This will be relevant once we extract the $D_{s0}(2317)$ dependence on the pion mass where we can understand that different charm quark masses are needed.

Given the large number of parameters, we apply the Least Absolute Shrinkage and Selection Operator (LASSO) \cite{Guegan:2015mea, lib1, lib2} method to eliminate the less relevant parameters of the model. In order to do that we will add a penalty term to the $\chi^2$ of the form:

\begin{equation}
    P = \frac{\lambda}{10} \sum_{i}^{n} |p_i|.
\end{equation}

As the penalty coefficient $\lambda$ increases, the LASSO method will set the less important parameters to zero in order to minimize the full penalized $\chi^2_F = \chi^2 + P$. 
The algorithm works as follows: We start by randomly dividing the original data set into five subsets. For each iteration we use four subsets for training the model by minimizing $\chi^2_F$ with the coordinate descent method. After that, the remaining subset will be used as test data set. We compute the $\chi^2$ with the parameters obtained from the training data. In our specific case, we have to add the data of the collaborations that only have one point (HSC \cite{cheungthomas} for light and heavy mass, S. Prelovsek \cite{Lang:2014yfa,aokiphys} and CLS \cite{brunomattia}) to the training data set in order to fit the model properly.

\begin{figure}[h!]
   \begin{minipage}{0.48\textwidth}
     \centering
     \includegraphics[width=0.9\linewidth]{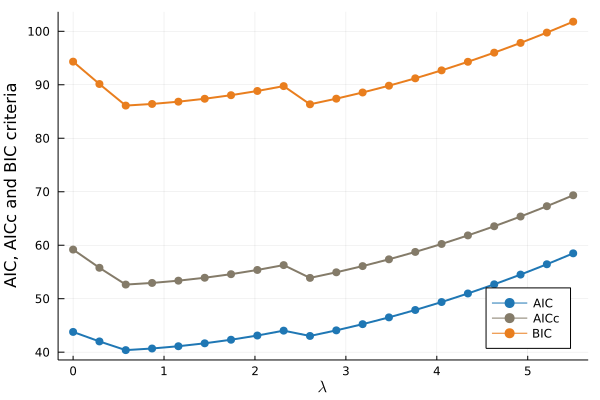}
   \end{minipage}
   \begin{minipage}{0.48\textwidth}
     \centering
     \includegraphics[width=0.9\linewidth]{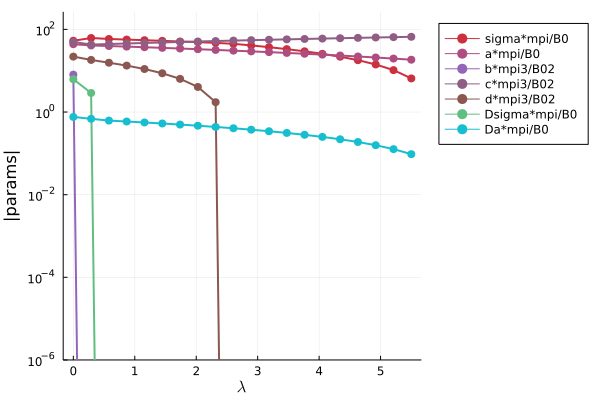}
   \end{minipage}
   \caption{LASSO method plots with lattice spacing error. Criteria information (left) and parameters evolution (right).}
   \label{lasso}
\end{figure}

\begin{figure}[h!]
   \begin{minipage}{0.48\textwidth}
     \centering
     \includegraphics[width=0.9\linewidth]{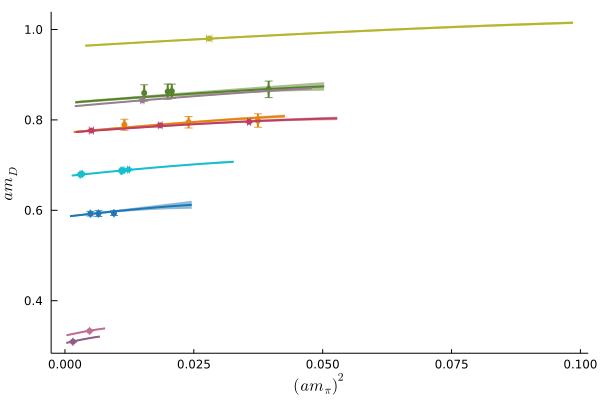}
   \end{minipage}
   \begin{minipage}{0.48\textwidth}
     \centering
     \includegraphics[width=0.9\linewidth]{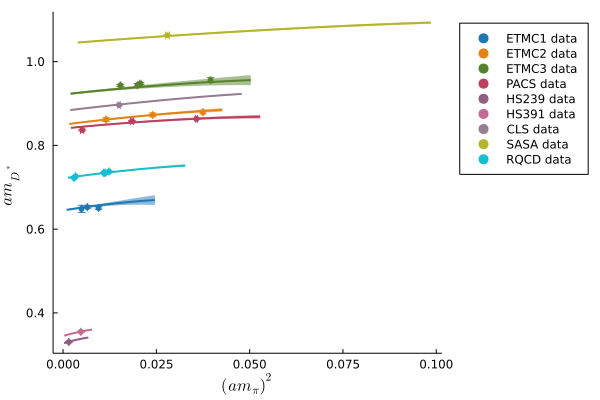}
   \end{minipage}
   \begin{minipage}{0.48\textwidth}
     \centering
     \includegraphics[width=0.9\linewidth]{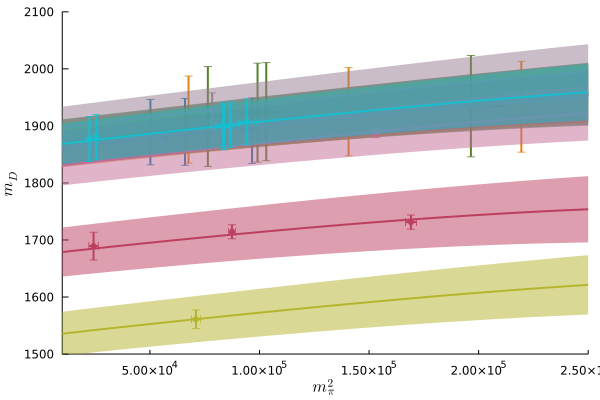}
   \end{minipage} 
   \begin{minipage}{0.48\textwidth}
     \centering
     \includegraphics[width=0.9\linewidth]{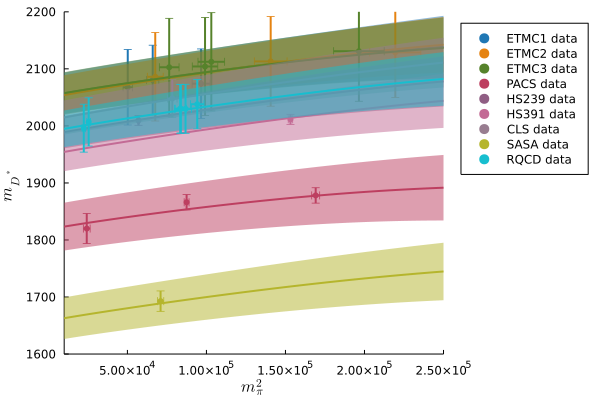}
   \end{minipage}
   \caption{After LASSO method fit without lattice spacing error (top) and with lattice spacing error (bottom).}
   \label{mDAF}
\end{figure}

We can see in FIG. \ref{lasso} that the information criteria shows minima around $\lambda =0.5$ and $\lambda =2.5$. This translates to eliminate two or three parameters. Then, we choose to eliminate only two, $bm^3_\pi/{B_0}^2$ and $\Delta^{(\sigma)}m_\pi/{B_0}$ because this choice minimizes the $\chi^2$ of the fit. Note that $bm^3_\pi/{B_0}^2$ corresponds to chiral lagrangian term with two insertions of the light quark mass matrix and $\Delta^{(\sigma)}m_\pi/{B_0}$ is a correction to the $\Delta$ parameter.
Then, we have reduced the degrees of freedom to $19$.
We use automatic differentiation \cite{Ramos:2018vgu} to evaluate the errors propagation with the Julia ADerrors package. We can see in FIG. \ref{mDAF} the final result of $D$ and $D^*$ meson masses fit to the lattice data, where in the bottom figure we have included also systematic errors for the lattice spacing.

Finally, if we want to extract results at the physical point, we need to obtain the parameters $m_H$ and $\Delta$ related to the physical charm quark mass and physical tree level hyperfine splitting. We fit these two extra parameters to the physical data from the PDG of the charmed mesons masses in order to obtain the right hierarchy.

\section{Energy levels fit}\label{Elvlsfit}

In order to extrapolate the $D_{s0}(2317)$ pole we need first to fit the indeterminate parameters of the $DK \rightarrow DK$ amplitude.
We fit to the new $N_f=2+1$ simulation that has been performed for $m_\pi=239$ MeV and $391$ MeV for $a_s=0.12$ fm and $L=1.9-3.8$ fm, where the $I=0$ $DK$ interaction is studied \cite{cheungthomas}.

\begin{figure}[h!]
   \begin{minipage}{0.48\textwidth}
     \centering
     \includegraphics[width=0.9\linewidth]{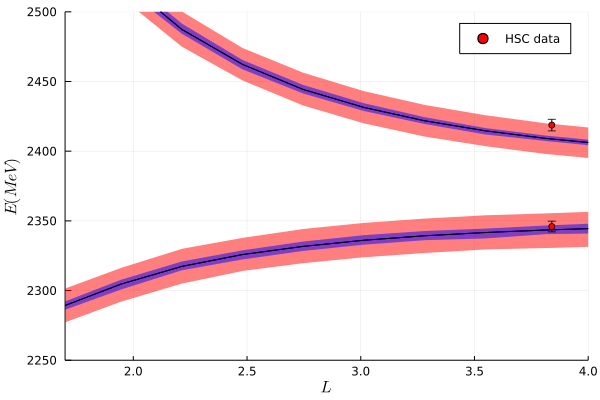}
   \end{minipage}
   \begin{minipage}{0.48\textwidth}
     \centering
     \includegraphics[width=0.9\linewidth]{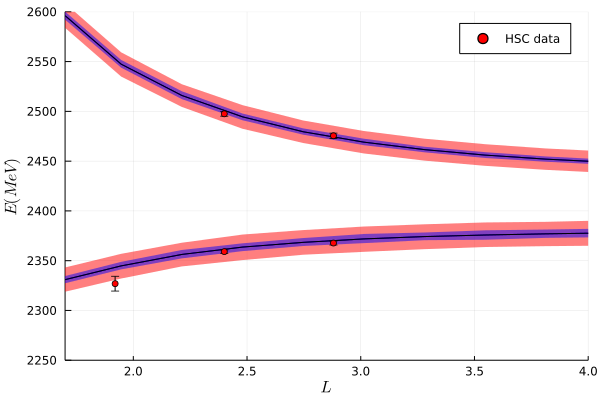}
   \end{minipage} 
   \caption{Results of the levels fit. We only show the energy levels corresponding to the symmetry $A_1^+$ and total momentum $P=(0,0,0)$. Red band is the errorbar taking into account the error of the lattice spacing and blue one with only the systematic error of the lattice data.}
   \label{Elevels}
\end{figure}

The scattering amplitude in the infinite volume is given by the Bethe-Salpeter equation, that is,
\begin{equation}
    T = \frac{V}{1-VG},
\end{equation}
where $V$ is the scattering potential matrix and $G$ the loop matrix.
To compute the $V$ potential for $DK \rightarrow DK$ we use the Hidden Gauge Lagrangian which corresponds to the hidden gauge formalism \cite{Molina:2010tx,Albaladejo:2018mhb,Gamermann:2006nm}, and also the lowest order of HM$\chi$PT. For the $DK \rightarrow DK$ scattering we only need the lagrangian term of pseudoscalar-pseudoscalar-vector that is, $\mathcal{L}^{PPV}=ig'Tr\left[[\partial_i\Phi,\Phi]\mathcal{V}^i\right]$ where $g'=m_{\rho}/(2f_{\pi})$, $\Phi$ is the $SU(4)$ pseudoscalar meson matrix given in \cite{Gamermann:2006nm} and $\mathcal{V}^i$ the vector meson matrix. With this, we have the following amplitude just before projecting to partial waves
\begin{equation}
    M_{DK\rightarrow DK} =-\frac{s-u}{2f^2}.
\end{equation}
To project to partial waves we use
\begin{equation}
    V_{l}(s)=\frac{1}{2}\int_{-1}^{1} M(s,\theta) \mathcal{P}_l(\cos(\theta)) d\cos(\theta),
\end{equation}
where $\mathcal{P}_l(x)$ are the Legendre polynomials.
The $G$ loop matrix elements has the form
\begin{equation}
    G^{(i)}_{inf}(E,q_{co}) = \int_0^{q_{co}} \frac{d^4q}{(2\pi)^4}\frac{1}{(P-q)^2-m_1^2+i\epsilon}\frac{1}{q^2-m_2^2+i\epsilon},
\end{equation}
where $m_i$ are the masses of the two mesons and $P$ the total momentum of the system. This function needs to be regularized and we can use three-momentum cutoff or dimensional regularization.
However, as we are going to fit to lattice data we need to discretize the momenta $q=(2\pi/L) \vec{n}$ with $\vec{n} \in \mathbb{Z}^3$. In the finite volume the scattering amplitude is given by $\widetilde{T}=V/(1-V\widetilde{G})$ and we can write $G$ matrix elements in the finite volume as
\begin{equation}
    G^{(i)}_{fin}(E,q_{co})=\frac{1}{L^3}\sum^{|\vec{q}|<q_{co}}_{\vec{n}} \frac{1}{2\omega_1(q)\omega_2(q)}\frac{\omega_1(q)\omega_2(q)}{E^2-(\omega_1(q)+\omega_2(q))^2+i\epsilon},
\end{equation}
where $\omega_i(q)=m_i^2+q^2$.
Following the same procedure than \cite{Doringg,MartinezTorres:2011pr} the final $\widetilde{G}$ that we insert to the Bethe-Salpeter equation is constructed with the elements
\begin{equation}\label{Gdef}
    \widetilde{G}^{(i)co}_{fin}= G^{(i)DR}_{inf}+ \lim_{q_{co}\rightarrow \infty}\left(G^{(i)}_{fin} - G^{(i)co}_{inf}\right).
\end{equation}
where superscripts $DR$ and $co$ means that we have used the dimensional and the cutoff regularization method respectively.
With this, the value of the cutoff is cancelled between the last two terms of Eq. (\ref{Gdef}). Fixing the dimensional regularization scale $\mu=1000$ MeV we only have the subtraction constant $a$ as a free parameter.
In order to extrapolate to the physical point and, since we have only HSC data for two different pion masses, we assume that the subtraction constant can be written as a first order taylor expansion of the squared pion mass as $a=\alpha  m_\pi^2 + \beta$. We determine the two fitting parameters $\alpha$, $\beta$ and, thus, $a|_{239}$ and $a|_{391}$.
In FIG. \ref{Elevels} we show the plot of the first energy levels for light and heavy pion mass as a function of the size of the volume $L$.

\begin{table}[h!]
\parbox{.45\linewidth}
{}
\centering
\begin{tabular}{|c|c|c|c|}\hline
$\alpha \cdot( 10^{-6} [\text{MeV}^{-2}])$ & $\beta$ & $a|_{239}$ & $a|_{391}$ \\ \hline
$-2.166$ & $-1.864$ & $-1.987$ & $-2.195$ \\ \hline
\end{tabular}
\caption{Parameters of the subtraction constant $a$ fit}
\label{aHfit}
\end{table}

\section{Pion mass dependence of the $D_{s0}(2317)$ and $D_{s1}(2460)$}\label{Ds0dependence}

\begin{figure}[h!]
   \begin{minipage}{0.48\textwidth}
     \centering
     \includegraphics[width=0.9\linewidth]{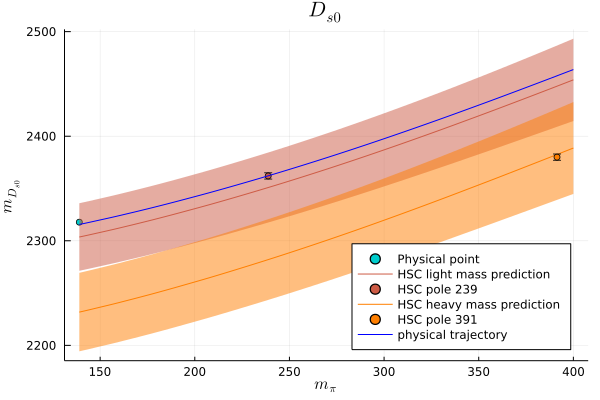}
   \end{minipage}
   \begin{minipage}{0.48\textwidth}
     \centering
     \includegraphics[width=0.9\linewidth]{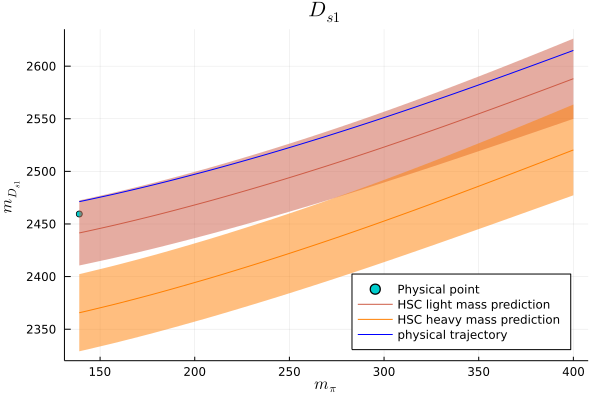}
   \end{minipage} 
   \caption{$D_{s0}(2317)$ and $D_{s1}(2460)$ masses dependence of the pion mass.}
   \label{Ds0andDs1}
\end{figure}

Finally, we have all we need to extract the dependence on the pion mass of the $D_{s0}(2317)$ resonance. Inserting the $D$ meson mass function that we have fitted previously in the Bethe-Salpeter equation for infinite volume, we can obtain the mass of the resonance by computing the poles of the equation for a given value of the pion mass. As we can compute the amplitude for the $D^*K \rightarrow D^*K$ process as well, we can also predict the $D_{s1}(2460)$ resonance pion mass dependence using the same procedure. We show these results in FIG. \ref{Ds0andDs1}.

We notice that, with the lattice spacings used in \cite{cheungthomas}, $a_t^{-1}=6079$ MeV for $m_\pi=239$ MeV and $a_t^{-1}=5667$ MeV for $m_\pi=391$ MeV, the charm quark mass determination of \cite{HadronSpectrum:2012gic} is lower for the heavy pion mass than for the light pion mass \cite{Cheung:2016bym} by around $20$ MeV, for that reason, two curves are plotted. The error bands are calculated by the error propagation of the fitted parameters from the $D^{(*)}$ meson masses analysis, which includes the lattice spacing systematic error.


The results that we have obtained are not straightforward at all. Just by looking at the two lattice data points of FIG. \ref{Ds0andDs1} (left) corresponding to the masses of the $D_{s0}(2317)$ obtained in \cite{cheungthomas} for the two different $a_tM_{\eta_c}$ determinations of Refs. \cite{HadronSpectrum:2012gic,Cheung:2016bym} one could think in principle that the $D_{s0}(2317)$ pole mass does not depend on the pion mass. But then, one would have to accept a systematic deviation of the lattice data from the experimental mass of the $D_{s0}(2317)$ of $\approx 30 - 40$ MeV. However, one needs to fit the energy levels of the $DK$ scattering and implement properly the light and heavy (charm) quark mass dependence, that is what we have done here. 

To summarize, our result supports a strong dependence of the $D_{s0}(2317)$ pole mass with the pion mass, and, thus, its molecular interpretation as $DK$ bound state. Our results also suggest that, given the evidence of many exotics observed very close to thresholds in the charm and hidden charm sectors, further improvement in controlling the systematic errors is needed in the lattice simulations to make a clear conclusion. For example, to evaluate the pole of the $D_{s0}(2317)$ at different pion masses for the same scale setting and charm quark mass, would be most welcome.

\section{Conclusions}

In this work we have extracted the parameters of one-loop HH$\chi$PT involved in the charmed mesons $D_{(s)}$ and $D_{(s)}^*$ formulas. With the help of the LASSO regression method we have eliminated superfluous degrees of freedom of the model. 
This has provided us with one tool to extrapolate the mass of the $D_{s0}(2317)$ to the physical point. After that, we have fitted the subtraction constants $a$ to the new lattice energy levels of HSC \cite{cheungthomas}. 
Finally, we could extract the pion mass dependence of the exotic resonances $D_{s0}(2317)$ and predict that for the $D_{s1}(2460)$.
We conclude that the pole of the $D_{s0}(2317)$ pion mass dependence does not agree with the assumption of a $c\bar{s}$ state, supporting its molecular interpretation. However, more precise lattice data for the same charm quark mass are needed to confirm this statement.

\section{Acknowledgments}

R. M. and F. Gil acknowledge support from the CIDEGENT program with Ref. CIDEGENT/2019/015 and from the spanish national grants PID2019-106080GB-C21 and PID2020-112777GB-I00. This project has received funding from the European Union’s Horizon 2020 programme No. 824093 for the STRONG-2020 project.



\bibliographystyle{JHEP}
\bibliography{biblio}       

\providecommand{\href}[2]{#2}\begingroup\raggedright\begin{thebibliography}{10}

\bibitem{guoreview}
F.-K.~Guo, C.~Hanhart, U.-G.~Mei\ss{}ner, Q.~Wang, Q.~Zhao and B.-S.~Zou,
  \emph{{Hadronic molecules}},
  \href{https://doi.org/10.1103/RevModPhys.90.015004}{\emph{Rev. Mod. Phys.}
  {\bfseries 90} (2018) 015004}
  [\href{https://arxiv.org/abs/1705.00141}{{\ttfamily 1705.00141}}].

\bibitem{babards0}
{\scshape BaBar} collaboration, \emph{{Observation of a narrow meson decaying
  to $D_s^+ \pi^0$ at a mass of 2.32-GeV/c$^2$}},
  \href{https://doi.org/10.1103/PhysRevLett.90.242001}{\emph{Phys. Rev. Lett.}
  {\bfseries 90} (2003) 242001}
  [\href{https://arxiv.org/abs/hep-ex/0304021}{{\ttfamily hep-ex/0304021}}].

\bibitem{cleods1}
{\scshape CLEO} collaboration, \emph{{Observation of a narrow resonance of mass
  2.46-GeV/c**2 decaying to D*+(s) pi0 and confirmation of the D*(sJ)(2317)
  state}}, \href{https://doi.org/10.1103/PhysRevD.68.032002}{\emph{Phys. Rev.
  D} {\bfseries 68} (2003) 032002}
  [\href{https://arxiv.org/abs/hep-ex/0305100}{{\ttfamily hep-ex/0305100}}].

\bibitem{godfrey}
S.~Godfrey and N.~Isgur, \emph{{Mesons in a Relativized Quark Model with
  Chromodynamics}}, \href{https://doi.org/10.1103/PhysRevD.32.189}{\emph{Phys.
  Rev. D} {\bfseries 32} (1985) 189}.

\bibitem{godfrey2}
S.~Godfrey and R.~Kokoski, \emph{{The Properties of p Wave Mesons with One
  Heavy Quark}}, \href{https://doi.org/10.1103/PhysRevD.43.1679}{\emph{Phys.
  Rev. D} {\bfseries 43} (1991) 1679}.

\bibitem{dipierro}
M.~Di~Pierro and E.~Eichten, \emph{{Excited Heavy - Light Systems and Hadronic
  Transitions}}, \href{https://doi.org/10.1103/PhysRevD.64.114004}{\emph{Phys.
  Rev. D} {\bfseries 64} (2001) 114004}
  [\href{https://arxiv.org/abs/hep-ph/0104208}{{\ttfamily hep-ph/0104208}}].

\bibitem{barnesclose}
T.~Barnes, F.E.~Close and H.J.~Lipkin, \emph{{Implications of a DK molecule at
  2.32-GeV}}, \href{https://doi.org/10.1103/PhysRevD.68.054006}{\emph{Phys.
  Rev. D} {\bfseries 68} (2003) 054006}
  [\href{https://arxiv.org/abs/hep-ph/0305025}{{\ttfamily hep-ph/0305025}}].

\bibitem{kolomeitsevlutz}
E.E.~Kolomeitsev and M.F.M.~Lutz, \emph{{On Heavy light meson resonances and
  chiral symmetry}},
  \href{https://doi.org/10.1016/j.physletb.2003.10.118}{\emph{Phys. Lett. B}
  {\bfseries 582} (2004) 39}
  [\href{https://arxiv.org/abs/hep-ph/0307133}{{\ttfamily hep-ph/0307133}}].

\bibitem{chenghou}
H.-Y.~Cheng and W.-S.~Hou, \emph{{B decays as spectroscope for charmed four
  quark states}},
  \href{https://doi.org/10.1016/S0370-2693(03)00834-7}{\emph{Phys. Lett. B}
  {\bfseries 566} (2003) 193}
  [\href{https://arxiv.org/abs/hep-ph/0305038}{{\ttfamily hep-ph/0305038}}].

\bibitem{terasaki}
K.~Terasaki, \emph{{BABAR resonance as a new window of hadron physics}},
  \href{https://doi.org/10.1103/PhysRevD.68.011501}{\emph{Phys. Rev. D}
  {\bfseries 68} (2003) 011501}
  [\href{https://arxiv.org/abs/hep-ph/0305213}{{\ttfamily hep-ph/0305213}}].

\bibitem{molinabranz}
R.~Molina, T.~Branz and E.~Oset, \emph{{A new interpretation for the
  $D^*_{s2}(2573)$ and the prediction of novel exotic charmed mesons}},
  \href{https://doi.org/10.1103/PhysRevD.82.014010}{\emph{Phys. Rev. D}
  {\bfseries 82} (2010) 014010}
  [\href{https://arxiv.org/abs/1005.0335}{{\ttfamily 1005.0335}}].

\bibitem{molinaoset}
R.~Molina and E.~Oset, \emph{{Molecular picture for the $X_0(2866)$ as a $D^*
  \bar{K}^*$ $J^P=0^+$ state and related $1^+,2^+$ states}},
  \href{https://doi.org/10.1016/j.physletb.2020.135870}{\emph{Phys. Lett. B}
  {\bfseries 811} (2020) 135870}
  [\href{https://arxiv.org/abs/2008.11171}{{\ttfamily 2008.11171}}].

\bibitem{LHCb1}
{\scshape LHCb} collaboration, \emph{{A model-independent study of resonant
  structure in $B^+\to D^+D^-K^+$ decays}},
  \href{https://doi.org/10.1103/PhysRevLett.125.242001}{\emph{Phys. Rev. Lett.}
  {\bfseries 125} (2020) 242001}
  [\href{https://arxiv.org/abs/2009.00025}{{\ttfamily 2009.00025}}].

\bibitem{LHCb2}
{\scshape LHCb} collaboration, \emph{{Amplitude analysis of the $B^+\to
  D^+D^-K^+$ decay}},
  \href{https://doi.org/10.1103/PhysRevD.102.112003}{\emph{Phys. Rev. D}
  {\bfseries 102} (2020) 112003}
  [\href{https://arxiv.org/abs/2009.00026}{{\ttfamily 2009.00026}}].

\bibitem{LHCb:2022xob}
{\scshape LHCb} collaboration, \emph{{First observation of a doubly charged
  tetraquark and its neutral partner}},
  \href{https://arxiv.org/abs/2212.02716}{{\ttfamily 2212.02716}}.

\bibitem{wise}
M.B.~Wise, \emph{{Chiral perturbation theory for hadrons containing a heavy
  quark}}, \href{https://doi.org/10.1103/PhysRevD.45.R2188}{\emph{Phys. Rev. D}
  {\bfseries 45} (1992) R2188}.

\bibitem{burdman}
G.~Burdman and J.F.~Donoghue, \emph{{Union of chiral and heavy quark
  symmetries}}, \href{https://doi.org/10.1016/0370-2693(92)90068-F}{\emph{Phys.
  Lett. B} {\bfseries 280} (1992) 287}.

\bibitem{yancheng}
T.-M.~Yan, H.-Y.~Cheng, C.-Y.~Cheung, G.-L.~Lin, Y.C.~Lin and H.-L.~Yu,
  \emph{{Heavy quark symmetry and chiral dynamics}},
  \href{https://doi.org/10.1103/PhysRevD.46.1148}{\emph{Phys. Rev. D}
  {\bfseries 46} (1992) 1148}.

\bibitem{jenkins}
E.E.~Jenkins, \emph{{Heavy meson masses in chiral perturbation theory with
  heavy quark symmetry}},
  \href{https://doi.org/10.1016/0550-3213(94)90499-5}{\emph{Nucl. Phys. B}
  {\bfseries 412} (1994) 181}
  [\href{https://arxiv.org/abs/hep-ph/9212295}{{\ttfamily hep-ph/9212295}}].

\bibitem{kalinowskiwagner}
M.~Kalinowski and M.~Wagner, \emph{{Masses of $D$ mesons, $D_s$ mesons and
  charmonium states from twisted mass lattice QCD}},
  \href{https://doi.org/10.1103/PhysRevD.92.094508}{\emph{Phys. Rev. D}
  {\bfseries 92} (2015) 094508}
  [\href{https://arxiv.org/abs/1509.02396}{{\ttfamily 1509.02396}}].

\bibitem{EuropeanTwistedMassa}
{\scshape European Twisted Mass} collaboration, \emph{{Up, down, strange and
  charm quark masses with N$_f$ = 2+1+1 twisted mass lattice QCD}},
  \href{https://doi.org/10.1016/j.nuclphysb.2014.07.025}{\emph{Nucl. Phys. B}
  {\bfseries 887} (2014) 19} [\href{https://arxiv.org/abs/1403.4504}{{\ttfamily
  1403.4504}}].

\bibitem{mohlerwoloshyn}
D.~Mohler and R.M.~Woloshyn, \emph{{$D$ and $D_s$ meson spectroscopy}},
  \href{https://doi.org/10.1103/PhysRevD.84.054505}{\emph{Phys. Rev. D}
  {\bfseries 84} (2011) 054505}
  [\href{https://arxiv.org/abs/1103.5506}{{\ttfamily 1103.5506}}].

\bibitem{aokiphys}
{\scshape PACS-CS} collaboration, \emph{{2+1 Flavor Lattice QCD toward the
  Physical Point}},
  \href{https://doi.org/10.1103/PhysRevD.79.034503}{\emph{Phys. Rev. D}
  {\bfseries 79} (2009) 034503}
  [\href{https://arxiv.org/abs/0807.1661}{{\ttfamily 0807.1661}}].

\bibitem{cheungohara}
{\scshape Hadron Spectrum} collaboration, \emph{{Excited and exotic charmonium,
  $D_s$ and $D$ meson spectra for two light quark masses from lattice QCD}},
  \href{https://doi.org/10.1007/JHEP12(2016)089}{\emph{JHEP} {\bfseries 12}
  (2016) 089} [\href{https://arxiv.org/abs/1610.01073}{{\ttfamily
  1610.01073}}].

\bibitem{cheungthomas}
{\scshape Hadron Spectrum} collaboration, \emph{{DK I = 0,$ D\overline{K} $I =
  0, 1 scattering and the $ {D}_{s0}^{\ast } $(2317) from lattice QCD}},
  \href{https://doi.org/10.1007/JHEP02(2021)100}{\emph{JHEP} {\bfseries 02}
  (2021) 100} [\href{https://arxiv.org/abs/2008.06432}{{\ttfamily
  2008.06432}}].

\bibitem{prelovsekpadmanath}
S.~Prelovsek, S.~Collins, D.~Mohler, M.~Padmanath and S.~Piemonte,
  \emph{{Charmonium-like resonances with J$^{PC}$ = 0$^{++}$, 2$^{++}$ in
  coupled $ \mathrm{D}\overline{\mathrm{D}} $, $
  {\mathrm{D}}_{\mathrm{s}}{\overline{\mathrm{D}}}_{\mathrm{s}} $ scattering on
  the lattice}}, \href{https://doi.org/10.1007/JHEP06(2021)035}{\emph{JHEP}
  {\bfseries 06} (2021) 035}
  [\href{https://arxiv.org/abs/2011.02542}{{\ttfamily 2011.02542}}].

\bibitem{brunomattia}
M.~Bruno, T.~Korzec and S.~Schaefer, \emph{{Setting the scale for the CLS $2 +
  1$ flavor ensembles}},
  \href{https://doi.org/10.1103/PhysRevD.95.074504}{\emph{Phys. Rev. D}
  {\bfseries 95} (2017) 074504}
  [\href{https://arxiv.org/abs/1608.08900}{{\ttfamily 1608.08900}}].

\bibitem{balicollins}
G.S.~Bali, S.~Collins, A.~Cox and A.~Sch\"afer, \emph{{Masses and decay
  constants of the $D_{s0}^*(2317)$ and $D_{s1}(2460)$ from $N_f=2$ lattice QCD
  close to the physical point}},
  \href{https://doi.org/10.1103/PhysRevD.96.074501}{\emph{Phys. Rev. D}
  {\bfseries 96} (2017) 074501}
  [\href{https://arxiv.org/abs/1706.01247}{{\ttfamily 1706.01247}}].

\bibitem{balicollinsa}
G.S.~Bali, S.~Collins, B.~Gl\"assle, M.~G\"ockeler, J.~Najjar, R.H.~R\"odl
  et~al., \emph{{Nucleon isovector couplings from $N_f=2$ lattice QCD}},
  \href{https://doi.org/10.1103/PhysRevD.91.054501}{\emph{Phys. Rev. D}
  {\bfseries 91} (2015) 054501}
  [\href{https://arxiv.org/abs/1412.7336}{{\ttfamily 1412.7336}}].

\bibitem{guoheo}
X.-Y.~Guo, Y.~Heo and M.F.M.~Lutz, \emph{{On chiral extrapolations of charmed
  meson masses and coupled-channel reaction dynamics}},
  \href{https://doi.org/10.1103/PhysRevD.98.014510}{\emph{Phys. Rev. D}
  {\bfseries 98} (2018) 014510}
  [\href{https://arxiv.org/abs/1801.10122}{{\ttfamily 1801.10122}}].

\bibitem{lutzsoyeur}
M.F.M.~Lutz and M.~Soyeur, \emph{{Radiative and isospin-violating decays of
  D(s)-mesons in the hadrogenesis conjecture}},
  \href{https://doi.org/10.1016/j.nuclphysa.2008.09.003}{\emph{Nucl. Phys. A}
  {\bfseries 813} (2008) 14} [\href{https://arxiv.org/abs/0710.1545}{{\ttfamily
  0710.1545}}].

\bibitem{Molina:2010tx}
R.~Molina, T.~Branz and E.~Oset, \emph{{A new interpretation for the
  $D^*_{s2}(2573)$ and the prediction of novel exotic charmed mesons}},
  \href{https://doi.org/10.1103/PhysRevD.82.014010}{\emph{Phys. Rev. D}
  {\bfseries 82} (2010) 014010}
  [\href{https://arxiv.org/abs/1005.0335}{{\ttfamily 1005.0335}}].

\bibitem{Albaladejo:2018mhb}
M.~Albaladejo, P.~Fernandez-Soler, J.~Nieves and P.G.~Ortega,
  \emph{{Contribution of constituent quark model $c\bar{s}$ states to the
  dynamics of the $D_{s0}^*(2317)$ and $D_{s1}(2460)$ resonances}},
  \href{https://doi.org/10.1140/epjc/s10052-018-6176-3}{\emph{Eur. Phys. J. C}
  {\bfseries 78} (2018) 722}
  [\href{https://arxiv.org/abs/1805.07104}{{\ttfamily 1805.07104}}].

\bibitem{Gamermann:2006nm}
D.~Gamermann, E.~Oset, D.~Strottman and M.J.~Vicente~Vacas, \emph{{Dynamically
  generated open and hidden charm meson systems}},
  \href{https://doi.org/10.1103/PhysRevD.76.074016}{\emph{Phys. Rev. D}
  {\bfseries 76} (2007) 074016}
  [\href{https://arxiv.org/abs/hep-ph/0612179}{{\ttfamily hep-ph/0612179}}].

\bibitem{Lutz:2022enz}
M.F.M.~Lutz, X.-Y.~Guo, Y.~Heo and C.L.~Korpa, \emph{{Coupled-channel dynamics
  with chiral long-range forces in the open-charm sector of QCD}},
  \href{https://arxiv.org/abs/2209.10601}{{\ttfamily 2209.10601}}.

\bibitem{MartinezTorres:2014kpc}
A.~Mart\'\i{}nez~Torres, E.~Oset, S.~Prelovsek and A.~Ramos, \emph{{Reanalysis
  of lattice QCD spectra leading to the $D_{s0}^*(2317)$ and
  $D_{s1}^*(2460)$}},
  \href{https://doi.org/10.1007/JHEP05(2015)153}{\emph{JHEP} {\bfseries 05}
  (2015) 153} [\href{https://arxiv.org/abs/1412.1706}{{\ttfamily 1412.1706}}].

\bibitem{Cheung:2016bym}
{\scshape Hadron Spectrum} collaboration, \emph{{Excited and exotic charmonium,
  $D_s$ and $D$ meson spectra for two light quark masses from lattice QCD}},
  \href{https://doi.org/10.1007/JHEP12(2016)089}{\emph{JHEP} {\bfseries 12}
  (2016) 089} [\href{https://arxiv.org/abs/1610.01073}{{\ttfamily
  1610.01073}}].

\bibitem{HadronSpectrum:2012gic}
{\scshape Hadron Spectrum} collaboration, \emph{{Excited and exotic charmonium
  spectroscopy from lattice QCD}},
  \href{https://doi.org/10.1007/JHEP07(2012)126}{\emph{JHEP} {\bfseries 07}
  (2012) 126} [\href{https://arxiv.org/abs/1204.5425}{{\ttfamily 1204.5425}}].

\bibitem{Guegan:2015mea}
B.~Guegan, J.~Hardin, J.~Stevens and M.~Williams, \emph{{Model selection for
  amplitude analysis}},
  \href{https://doi.org/10.1088/1748-0221/10/09/P09002}{\emph{JINST} {\bfseries
  10} (2015) P09002} [\href{https://arxiv.org/abs/1505.05133}{{\ttfamily
  1505.05133}}].

\bibitem{lib1}
T.~Hasti, R.~Tibshirani and J.~Friedman, \emph{{The Elements of Statistical
  Learning: Data Mining, Inference, and Prediction}}, {\emph{Springer-Verlag}
  (2009) }.

\bibitem{lib2}
G.~James, D.~Witten, T.~Hastie and R.~Tibshirani, \emph{{An Introduction to
  Statistical Learning}}, {\emph{Springer-Verlag} (2013) }.

\bibitem{Lang:2014yfa}
C.B.~Lang, L.~Leskovec, D.~Mohler, S.~Prelovsek and R.M.~Woloshyn, \emph{{Ds
  mesons with DK and D*K scattering near threshold}},
  \href{https://doi.org/10.1103/PhysRevD.90.034510}{\emph{Phys. Rev. D}
  {\bfseries 90} (2014) 034510}
  [\href{https://arxiv.org/abs/1403.8103}{{\ttfamily 1403.8103}}].

\bibitem{Ramos:2018vgu}
A.~Ramos, \emph{{Automatic differentiation for error analysis of Monte Carlo
  data}}, \href{https://doi.org/10.1016/j.cpc.2018.12.020}{\emph{Comput. Phys.
  Commun.} {\bfseries 238} (2019) 19}
  [\href{https://arxiv.org/abs/1809.01289}{{\ttfamily 1809.01289}}].

\bibitem{Doringg}
M.~Doring, U.G.~Meissner, E.~Oset and A.~Rusetsky, \emph{{Scalar mesons moving
  in a finite volume and the role of partial wave mixing}},
  \href{https://doi.org/10.1140/epja/i2012-12114-6}{\emph{Eur. Phys. J. A}
  {\bfseries 48} (2012) 114} [\href{https://arxiv.org/abs/1205.4838}{{\ttfamily
  1205.4838}}].

\bibitem{MartinezTorres:2011pr}
A.~Martinez~Torres, L.R.~Dai, C.~Koren, D.~Jido and E.~Oset, \emph{{The $KD$,
  $\eta D_s$ interaction in finite volume and the nature of the $D_{s^*
  0}(2317)$ resonance}},
  \href{https://doi.org/10.1103/PhysRevD.85.014027}{\emph{Phys. Rev. D}
  {\bfseries 85} (2012) 014027}
  [\href{https://arxiv.org/abs/1109.0396}{{\ttfamily 1109.0396}}].

\end{thebibliography}\endgroup



\providecommand{\href}[2]{#2}\begingroup\raggedright\endgroup


\end{document}